\documentclass[aps,pra,preprint,superscriptaddress,longbibliography]{revtex4-2}

\usepackage{amsmath}
\usepackage{amsthm}
\usepackage{amsfonts}
\usepackage{amssymb}
\usepackage{xcolor,graphicx}
\usepackage{tikz}
\usepackage{color}
\usepackage[pdftex]{hyperref} 
\usepackage{longtable}
\usepackage{array}
\usepackage{dsfont}

\begin{document}
	
	\title{Continuous-time quantum harmonic oscillator state engineering}

	\author{E. Garc\'ia Herrera}
	\email[e-mail: ]{a01274795@tec.mx}
	\affiliation{Tecnologico de Monterrey, Escuela de Ingenier\'ia y Ciencias, Ave. Eugenio Garza Sada 2501, Monterrey, N.L., Mexico, 64849}

	\author{F. Torres-Leal}
	\email[e-mail: ]{a01067732@tec.mx}
	\affiliation{Tecnologico de Monterrey, Escuela de Ingenier\'ia y Ciencias, Ave. Eugenio Garza Sada 2501, Monterrey, N.L., Mexico, 64849} 	
	
	\author{B. M. Rodr\'iguez-Lara}
    \email[e-mail: ]{bmlara@tec.mx}
	\affiliation{Tecnologico de Monterrey, Escuela de Ingenier\'ia y Ciencias, Ave. Eugenio Garza Sada 2501, Monterrey, N.L., Mexico, 64849}	
	
	\date{\today}
	
	\begin{abstract}
    The center of mass motion of trapped ions and neutral atoms is suitable for approximation by a time-dependent driven quantum harmonic oscillator whose frequency and driving strength may be controlled with high precision.
    We show the time evolution for these systems with continuous differentiable time-dependent parameters in terms of the three basic operations provided by its underlying symmetry, rotation, displacement, and squeezing, using a Lie algebraic approach. 
    Our factorization of the dynamics allows for the intuitive construction of protocols for state engineering, for example, creating and removing displacement and squeezing, as well as their combinations, optimizing squeezing, or more complex protocols that work for slow and fast rates of change in the oscillator parameters. 
   \end{abstract}
	
	
	\maketitle
	\newpage

\section{Introduction}

The center of mass motion of ions in Paul traps \cite{Brown1991p527, Leibfried2003p281, Mihalcea2009p014006} and of neutral atoms in optical lattices \cite{Morinaga1999p4037, Calarco2000p022304, Creffield2007p110501} may be approximated by a time-dependent driven quantum harmonic oscillator (QHO). 
The parameters of the oscillator, such as its resonant frequency and external driving strength, may be controlled with high precision \cite{Zhang2006p042316, Forster2009p233001}. 
This allows for the manipulation of the center of mass motion of these trapped particles for various applications. 

In quantum information theory with continuous variables \cite{Weedbrook2012p621}, these systems may be used for Gaussian boson sampling \cite{Thekkadath2022p020336} or quantum error correction using squeezed Schr\"odinger cat states \cite{Schlegel2022p022431}. 
They may be used for the transport, separation, and merging of trapped ions in quantum computing architectures using sequences of time-varying potentials that produce squeezing and displacement of their center of mass motion state \cite{Sutherland2021p083201}.
In the fabrication of quantum technologies platforms, there exists an interesting theoretical proposal for the application of continuous-time-dependent systems in the generation of nitrogen ion beams with well-defined momentum for the creation of nitrogen-vacancy color centers localized at well-defined penetration depths in diamonds \cite{Tobalina2022p2239} where current approaches utilize invariants of the dynamics \cite{Tobalina2020p063112, Muga2020p043162} to propose shortcuts to adiabaticity. 
It may be possible to use phase-coherent sensing of the center-of-mass motion of trapped ions for weak force detection in fundamental physics \cite{Affolter2020p052609}.

It is well-known that sudden frequency jumps in the quantum harmonic oscillator produce squeezing \cite{Janszky1986p151, Graham1987p873, Brown1991p527, Janszky1992p6091}.
There exist theoretical protocols for squeezing maximization using optimized sudden frequency jumps providing an exponential increase in the squeezing of the motional state of trapped ions \cite{Galve2009p055804}.
In the experimental front, demonstrations of this effect for the quantum motion of the center of mass of an ion in a Paul trap \cite{Leibfried2003p281} or of neutral cold atoms trapped in an optical lattice \cite{Xin2021p183602} exist.
However, it was recently shown that the speed of change in the frequency and its direction affects the squeezing \cite{MartinezTibaduiza2021p205401}; that is, adiabatic changes produce no squeezing, while slow changes may produce more complex states. 

We are interested in the time-dependent driven QHO from a Lie theory point of view that provides a framework to optimize state engineering at any given rate of change.
In Sec. \ref{sec:Model}, we present the model, its underlying Lie algebra, and the associated Lie group defining the three basic operations provided by the dynamics; that is, rotation, displacement, and squeezing.
These unitary operations help us construct the dynamics of any given initial state under the evolution of the driven QHO with continuous and differentiable time-dependent parameters in Sec. \ref{sec:Dynamics}.
The intuition gained by the shape of the final state as a series of rotations, displacement and squeezing of the initial state helps us provide examples of continuous-time state engineering in Sec. \ref{sec:Examples}.
We close with our conclusion in Sec. \ref{sec:Conclusion}.

\section{Model and its symmetries} \label{sec:Model}

We are interested in a time-dependent Hamiltonian model that may describe the quantum motion of the center of mass of an ion in a Paul trap or of neutral cold atoms trapped in an optical lattice under external driving,
\begin{align}\label{eq:Model}
    \hat{H}(t) = \frac{1}{2 m} \hat {p}^{2} + \frac{1}{2} m \omega^{2}(t) \hat{q}^{2} + \Omega(t) \cos \left( \omega_{d} t + \phi \right) \hat{q},
\end{align}
where the mass $m$ is a real positive constant, the resonant frequency of the oscillator $\omega(t)$ is a real, positive definite, smooth, and continuous function of time, the external driving strength $\Omega(t)$ is a real, smooth, and continuous function of time, and the external driving frequency $\omega_{d}$ is a real positive constant.
This model belongs to a Hamiltonian class with a five-element underlying Lie algebra \cite{Onah2022p0000},
\begin{equation}
\begin{aligned}
    \left[\hat{q},\hat{p}\right]&=i\hbar,&\left[\{\hat{q},\hat{p}\},\hat{q}\right]&=-2i\hbar\hat{q},\\
    \left[\hat{q},\hat{p}^{2}\right]&=2i\hbar\hat{p},&\left[\{\hat{q},\hat{p}\},\hat{p}\right]&=2i\hbar\hat{p},\\
    \left[\hat{p},\hat{q}^{2}\right]&=-2i\hbar\hat{q},&\left[\{\hat{q},\hat{p}\},\hat{q}^{2}\right]&=-4i\hbar\hat{q}^{2}, \\
    \left[\hat{q}^{2},\hat{p}^{2}\right]&=2i\hbar\{\hat{q},\hat{p}\},&\left[\{\hat{q},\hat{p}\},\hat{p}^{2}\right]&=4i\hbar\hat{p}^{2}, \\
\end{aligned}  
\end{equation}
defining three types of unitary transformations.
A two-parameter translation or displacement, 
\begin{align}
    \hat{D}\left( \beta_{q}, \beta_{p} \right) =&~ e^{- \frac{i}{\hbar} \left[ \beta_{p} \hat{q} + \beta_{q} \hat{p} \right]},   
\end{align}
in terms of the real parameters $\beta_{q}$ and $\beta_{p}$ with units of $\mathrm{position}$ and $\mathrm{momentum}$, in that order, such that the canonical position and momentum are displaced by its corresponding parameter,
\begin{align}
    \hat{D}^{\dagger}\left( \beta_{q}, \beta_{p} \right) \, \hat{q} \,  \hat{D}\left( \beta_{q}, \beta_{p} \right) & = \hat{q} + \beta_{q}, \\
    \hat{D}^{\dagger}\left( \beta_{q}, \beta_{p} \right) \, \hat{p} \, \hat{D}\left( \beta_{q}, \beta_{p} \right) &= \hat{p} - \beta_{p}.
\end{align}
A two-parameter rotation with scaling, 
\begin{align}
    \hat{R}(\theta_{q}, \theta_{p}) =&~ e^{- \frac{i}{\hbar} \left[ \theta_{q}^{2} \hat{q}^{2} + \theta_{p}^{2} \hat{p}^{2} \right]}, 
\end{align}
with real parameters $\theta_{q}$ and $\theta_{p}$ with units of $\mathrm{action}^{1/2} \cdot \mathrm{position}^{-1}$ and $\mathrm{action}^{1/2} \cdot \mathrm{momentum}^{-1}$, in that order, such that they provide a rotation proportional to the product of the parameters and scaling in the canonical pair proportional to their quotient,
\begin{align}
    \hat{R}^{\dagger}(\theta_{q}, \theta_{p}) \, \hat{q} \, \hat{R}(\theta_{q}, \theta_{p}) &= \cos{\left(2 \theta_{q} \theta_{p} \right)} \, \hat{q} + \frac{ \theta_{p}}{\theta_{q} } \sin{\left(2 \theta_{q}  \theta_{p} \right)} \,  \hat{p}, \\
    \hat{R}^{\dagger}(\theta_{q}, \theta_{p}) \, \hat{p} \, \hat{R}(\theta_{q}, \theta_{p}) &= \cos{\left(2  \theta_{q}  \theta_{p} \right)} \, \hat{p} - \frac{\theta_{q}}{ \theta_{p}} \sin{\left(2 \theta_{q}  \theta_p \right)} \,  \hat{q},
    \label{rotation}
\end{align}
for example, the standard time-independent harmonic oscillator provides an evolution operator of this form with $2 \theta_{q} \theta_{p} = \omega t$ and $\theta_{q} / \theta_{p} = m \omega$, leading to circular orbits for expected values of the canonical pair around the origin of phase space.
Finally, the symmetries of the model provide us with a squeezing,
\begin{align}
    \hat{S}(r) =&~ e^{- \frac{i}{\hbar} r \left\{ \hat{q}, \hat{p} \right\}},
\end{align}
in terms of the real dimensionless parameter $r$ that provides exponential scaling for the canonical pair, 
\begin{align}
    \hat{S}^{\dagger}(r) \, \hat{q} \, \hat{S}(r) &= \hat{q} e^{2 r}, \\
    \hat{S}^{\dagger}(r) \, \hat{p} \, \hat{S}(r) &= \hat{p} e^{ -2 r }.
\end{align} 
These actions of the elements of the group on the elements of the algebra provide us with the theoretical tools to understand its dynamics.

\section{Dynamics} \label{sec:Dynamics}

The three unitary operations of the Lie group associated with the underlying Lie algebra of our model allow us to discuss its dynamics in various reference frames.
Let us start our analysis of the dynamics with a time-dependent displacement, 
\begin{align} \label{Eliminate_x_p_terms}
    \vert {\psi} (t) \rangle = \hat{D}\left[ \beta_{q}(t), \beta_{p}(t) \right] \vert {\psi_{1}} (t) \rangle,
\end{align}
where the real parameters $\beta_{q}(t)$ and $\beta_{p}(t)$ answer to the differential equation set,
\begin{align}
    \ddot{\beta}_{q}(t) + \omega^{2}(t) \beta_{q}(t) =& - \frac{\Omega (t)}{m} \cos{\left( \omega_d t + \phi \right)}, \\
    \beta_p (t) =& -m \dot{\beta}_{q} (t),
\end{align}
where we use the shorthand notation $\dot{f} \equiv d f / dt$ and $\ddot{f} \equiv d^{2} f / dt^{2}$ for the time derivatives.
The displacement parameters fulfill the  initial conditions, 
\begin{align}
    \beta_{q}(0) =& - \frac{\Omega(0) \cos{\left( \phi \right)}}{m \omega^2(0)}, \\
    \beta_{p}(0) =& 0,
\end{align}
in terms of the initial values for the resonant frequency and external driving strength.
This time-dependent displacement allows us to rewrite our model Hamiltonian without the external drive,
\begin{align}
\hat{H}_{1}(t) = \frac{1}{2 m} \hat {p}^{2} + \frac{1}{2} m \omega^{2}(t) \hat{q}^{2} + \ell(t),
\end{align}
plus a time-dependent energy bias term that looks like the energy of a classical forced harmonic oscillator,
\begin{align}
    \ell (t) =&~ \frac{1}{2m} \beta_{p}^{2}(t) + \frac{1}{2} m \omega^{2} (t) \beta_{q}^2 (t) + \Omega (t) \cos{\left(\omega_{d} t + \phi \right)} \beta_{q} (t),
\end{align}
where the parameters $\beta_{q}$ and $\beta_{p}$ play the role of position and momentum.  
In order to remove this energy bias, we move to a reference frame provided by a time-dependent overall phase,
\begin{align} \label{Eliminate_bias_term}
    \vert {\psi_{1}} (t)\rangle = e^{ - \frac{ i }{ \hbar } \int_0^{t} \ell( \tau ) d \tau } \vert{\psi_{2}} (t)\rangle,
\end{align}
that yields an effective Paul-trap Hamiltonian,
\begin{align}
\hat{H}_{2}(t) = \frac{1}{2 m} \hat {p}^{2} + \frac{1}{2} m \omega^{2}(t) \hat{q}^{2},
\end{align}
with a time-dependent resonant frequency. 
It is feasible to rewrite this effective time-dependent Hamiltonian into one where the time-dependence is factored out using a squeezing and momentum displacement transformation \cite{BMRodriguez2014p2083},
\begin{align}\label{Factorize_time_dependence}
    \vert \psi_{2}(t) \rangle = \hat{R}\left[ \theta_{q}(t), 0 \right] \hat{S}\left[ r(t) \right]  \vert{\psi_{3}}(t) \rangle,
\end{align}
where the parameters for the rotation and the squeezing, 
\begin{align}
    \theta_{q}(t) &= - \frac {m}{2} \frac { \dot {\rho} (t)}{ \rho (t)}, \\
    r(t) &= \frac {1}{2} \ln \left[\sqrt{m \omega (0)}\rho \left(t \right) \right],
\end{align}
are given in terms of the auxiliary real function $\rho(t)$ fulfilling the Ermakov equation,
\begin{align}
    \ddot {\rho} (t) + \omega^{2} (t) \rho (t) = \frac {1} {m^{2} \rho^{3} (t)}.
\end{align}
and the initial conditions,
\begin{align}
    \rho(0) =& \frac{1}{\sqrt{m \omega(0)}},\\
    \dot{\rho}(0) =& 0,
\end{align}
providing us with the identity operator at the initial time $t=0$.
The effective Hamiltonian in this reference frame, 
\begin{align}
    \hat{H}_{3}(t) = \frac{1}{m \omega(0) \rho^{2}(t)} \left[ \frac{1}{2m} \hat{p}^{2} + \frac{1}{2} m \omega^2(0) \hat{q}^2  \right],
\end{align}
is composed of a time-dependent scalar factor multiplying a time-independent quantum harmonic oscillator with a resonant frequency equal to that at the initial time $\omega(0)$.
Thus, it is possible to write the time evolution in this frame, 
\begin{align}
    \vert \psi_{3}(t) \rangle = \hat{R}\left[ \varphi_{q}(t), \varphi_{p}(t) \right] \vert \psi_{3}(0) \rangle,
\end{align}
as a rotation with real parameters,
\begin{align}
    \varphi_{q}^{2}(t) &= \frac{1}{2} \omega(0) \int_{0}^{t} \frac{1}{\rho^{2}(\tau)} d\tau, \\
    \varphi_{p}^{2}(t) &= \frac{1}{2m^{2} \omega(0)}  \int_{0}^{t} \frac{1}{\rho^{2}(\tau)} d\tau,
\end{align}
in terms of the initial value of the resonant frequency and the auxiliary Ermakov function.

In consequence, we may express the evolution of the initial state of the system, 
\begin{align}
    \vert \Psi(t) \rangle =&~  e^{ - \frac{ i }{ \hbar } \int_0^{t} \ell( \tau ) d \tau }  \hat{D}\left[ \beta_{q}(t), \beta_{p}(t) \right] \hat{R}\left[ \theta_{q}(t), 0 \right]  \hat{S}\left[ r(t) \right] \hat{R}\left[ \varphi_{q}(t), \varphi_{p}(t) \right] \times \\
    &~ \times \hat{D}\left[ \frac{\Omega(0) \cos{\left( \phi \right)}}{m \omega^2(0)}, 0 \right] \vert \Psi(0) \rangle, 
\end{align}
as a sequence of the three basic operations, that is displacement, rotation, and squeezing, provided by underlying symmetries of the model.
The initial state of the system is displaced in canonical position by a factor proportional to the driving strength at the initial time and inversely proportional to the mass and initial resonant frequency that does not change the shape of its quasi-probability distribution in phase space. 
Then, the rotation of the canonical pair makes the quasi-probability distribution spin over the axis defined by the expected value of the canonical pair and circle the origin of phase space itself with the frequency defined by the time-dependent phase $ 2 \varphi_{q}(t) \varphi_{q}(t) = m^{-1} \int_{0}^{t} \rho^{-2}(\tau) d\tau$ as the scaling is time-independent, $\varphi_{q}(t) / \varphi_{p}(t) = m \omega (0)$ without changing its general shape.
Subsequently, the state is squeezed proportional to the logarithm of the auxiliary Ermakov parameter with the quasi-probability distribution in phase space expands (compresses) along the position (momentum) axis.
The following rotation is interesting because it does not affect the canonical position, it displaces the canonical momentum by a factor proportional to the canonical position as $\lim_{\theta_{p} \rightarrow 0} \theta_{q}(t) \sin (2 \theta_{q}(t) \theta_{p} ) / \theta_{p} = 2 \theta_{q}^{2}(t)$, and changes the shape of the quasi-probability distribution in phase space.
Then, a time-dependent displacement in the canonical pair is followed by an overall time-dependent phase that has no effect on measurements or the quasi-probability distribution.
We want to stress that, depending on the functional form of the resonant frequency and driving strength, we have access to all displacement, rotation, and squeezing parameters, at least numerically in the worst-case scenario, in order to follow the dynamics of the system.
 
\section{Continuous-time state engineering} \label{sec:Examples}

Our analytic description delivers some straightforward processes for quantum state engineering. 
For the sake of simplicity and to create some insight, let us consider that the pump always takes a null value at initial time, $\Omega(0) = 0$, the initial frequency of the trap, $\omega(0) = \omega_{0}$, serves as our reference frequency, and the general initial state  $\vert \Psi(0) \rangle $, such that the state of the system is the following,
\begin{align}
    \vert \Psi(t) \rangle =&~  e^{ - \frac{ i }{ \hbar } \int_0^{t} \ell( \tau ) d \tau }  \hat{D}\left[ \beta_{q}(t), \beta_{p}(t) \right] \hat{R}\left[ \theta_{q}(t), 0 \right]  \hat{S}\left[ r(t) \right] \hat{R}\left[ \varphi_{q}(t), \varphi_{p}(t) \right] \vert \Psi(0) \rangle, 
\end{align}
where we may neglect the overall phase and focus on the three unitary transformations.

\begin{figure}
    \centering
    \includegraphics{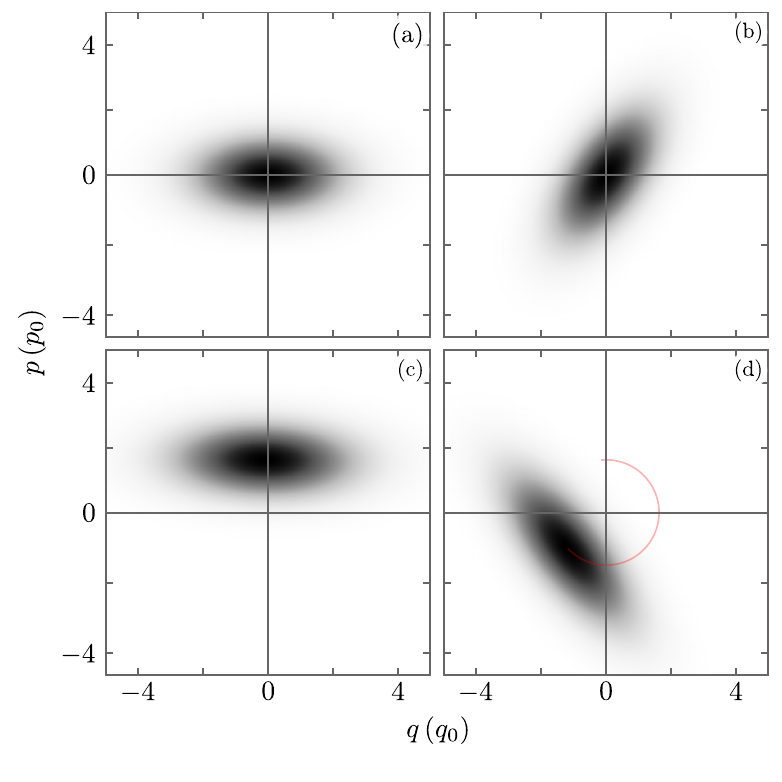}
    \caption{Husimi Q-function for the state of a system under QHO dynamics with constant resonant frequency $\omega_{0}$. (a) Initial squeezed state at dimensionless time $\omega_{0} t = 0$ and (b) final state at $\omega_{0} t = \pi/4$ and (c) initial displaced squeezed state at $\omega_{0} t = 0$ and (d) final state at $\omega_{0} t = 3 \pi/4$ where we show the expectation values of the canonical pair in red.}
    \label{fig:Figure1}
\end{figure}

\begin{figure}
    \centering
    \includegraphics{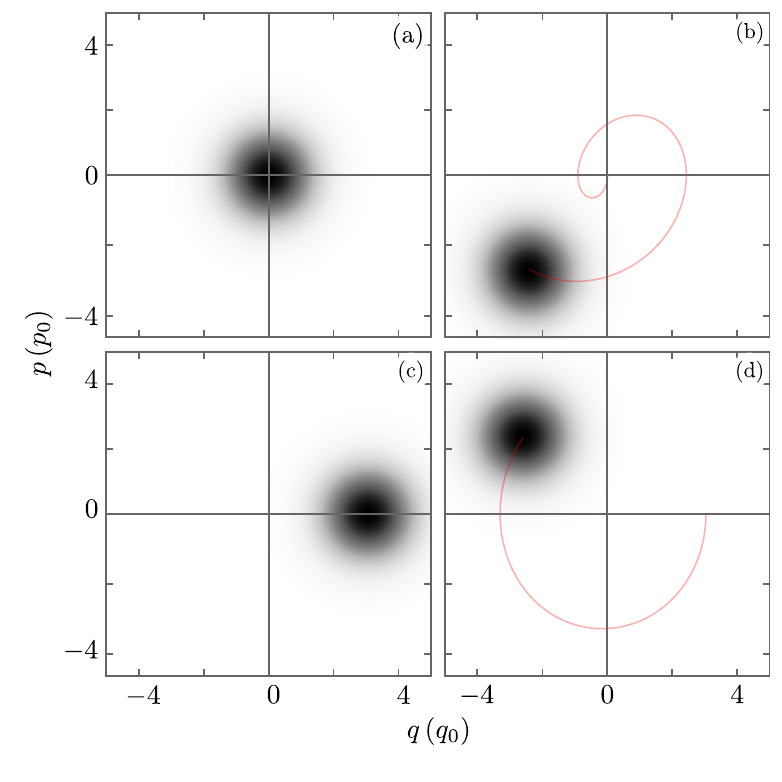}
    \caption{Husimi Q-function for the state of a system under driven QHO dynamics with constant resonant frequency $\omega_{0}$ and driving strength $\Omega_{0}$ . (a) Initial ground state at dimensionless time $\omega_{0} t = 0$ and (b) final state at $\omega_{0} t = 5 \pi/4$ and (c) initial displaced ground state at $\omega_{0} t = 0$ and (d) final state at $\omega_{0} t = 3 \pi/4$ where we show the expectation values of the canonical pair in red.}
    \label{fig:Figure2}
\end{figure}

\begin{figure}
    \centering
    \includegraphics{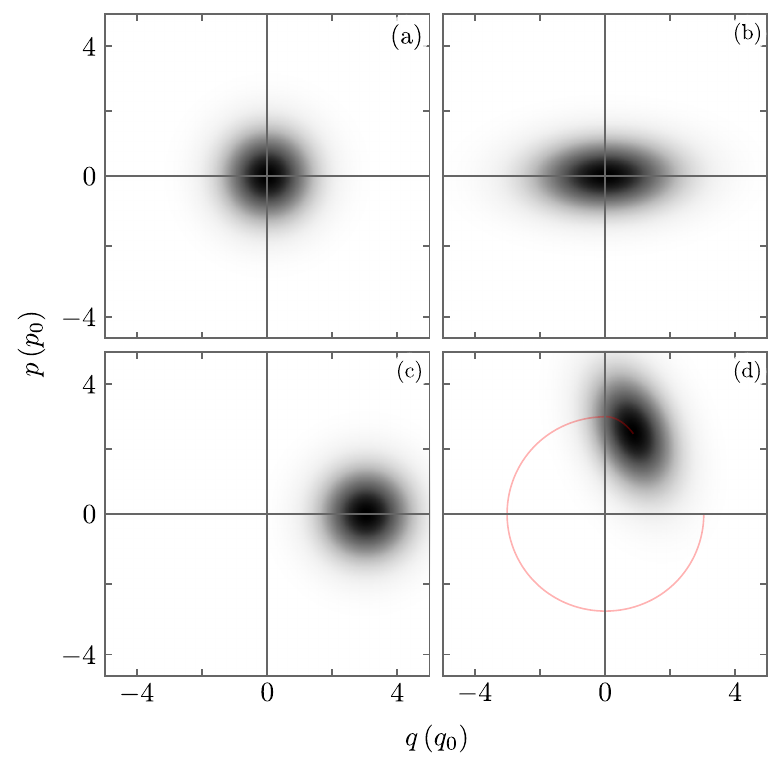}
    \caption{Husimi Q-function for the state of a system under QHO dynamics with time-dependent frequency $\omega(t)$. (a) Initial ground state at dimensionless time $\omega(0) t = 0$ and (b) final state at $\omega(0) t = \pi/2$ and (c) initial displaced ground state at $\omega(0) t = 0$ and (d) final state at $\omega(0) t = 5 \pi/4$ where we show the expectation values of the canonical pair in red.}
    \label{fig:Figure3}
\end{figure}

In the simplest case, a time-independent quantum harmonic oscillator, the state of the system,
\begin{align}
    \vert \Psi(t) \rangle =&~ \hat{R}\left[ \frac{1}{2} m \omega_{0}^{2} t, \frac{t}{2m}  \right] \vert \Psi(0) \rangle, 
\end{align}
for any given initial state, leads to a circular trajectory of the center of its quasi-probability distribution, for example  Husimi Q-function, 
\begin{align}
    Q(q,p) = \frac{1}{\pi} \langle \alpha \vert \hat{\rho} \vert \alpha \rangle,
\end{align}
with coherent parameter, $ \alpha = ( q/q_{0} + i p/p_{0} ) / \sqrt{2} $ and harmonic oscillator re-normalization parameters $q_{0} = \sqrt{\hbar / \left( m \omega_{0} \right)}$ and $p_{0} = \sqrt{\hbar m \omega_{0}}$, around the origin of phase space with frequency $\omega_{0}$ and a rotation of the quasi-probability distribution about its center, Fig. \ref{fig:Figure1}, due to the expected values of the canonical pair taking the form,
\begin{align} 
  \langle \hat{q}(t) \rangle &=  \cos{\left( \omega_{0} t \right)}  \langle \hat{q}(0) \rangle + \frac{1}{m \omega_{0}} \sin\left( \omega_{0} t \right) \, \langle \hat{p}(0) \rangle, \\
    \langle \hat{p}(t) \rangle &=\cos{\left(\omega_{0} t \right)} \, \langle \hat{p}(0) \rangle - m \omega_{0} \sin{\left( \omega_{0} t \right)} \,  \langle \hat{q}(0) \rangle,
\end{align}
where we used the shorthand notation $\langle \hat{o}(t) \rangle = \mathrm{Tr} \left[ \hat{\rho}(t) \hat{o} \right]$ for the expectation value of an operator.
Adding driving, the final state of the system is a displaced state, 
\begin{align} 
    \vert \Psi(t) \rangle =&~  \hat{D}\left[ \beta_{q}(t), \beta_{p}(t) \right]  \hat{R}\left[ \varphi_{q}(t), \varphi_{p}(t) \right] \vert \Psi(0) \rangle, 
\end{align}
with expectation values for the canonical pair, 
\begin{align} \label{eq:expval_displacement}
  \langle \hat{q}(t) \rangle &=   \cos{\left( \omega_{0} t \right)}  \, \langle \hat{q}(0) \rangle  + \frac{1}{m \omega_{0}} \sin\left( \omega_{0} t \right) \, \langle \hat{p}(0) \rangle  + \beta_{q}(t) , \\
    \langle \hat{p}(t) \rangle &=\cos{\left(\omega_{0} t \right)} \, \langle \hat{p}(0) \rangle - m \omega_{0} \sin{\left( \omega_{0} t \right)} \, \langle \hat{q}(0) \rangle - \beta_{p}(t),
\end{align}
that define a trajectory for the center of the quasi-probability rotating about the origin of phase space and its center with frequency $\omega_{0}$ plus following the parametric trajectory defined by $\{ \beta_{q}(t), - \beta_{p}(t) \}$. 
If the initial state is the ground state of the harmonic oscillator, the final state is a coherent state that follows the parametric trajectory $\{ \beta_{q}(t), - \beta_{p}(t) \}$,  Fig. \ref{fig:Figure2}.
The third simple case consists of a time-dependent frequency with no driving, provides a final state
\begin{equation}
    \vert{\psi (t)} \rangle = \hat{R}\left[\theta_{q} (t), 0 \right] \hat{S}\left[r (t) \right] \hat{R} \left[ \varphi_{q} (t), \varphi_{p} (t) \right] \vert{\psi (0)} \rangle,
\end{equation}
that is rotated by the standard quantum harmonic oscillator with time-dependent parameters, followed by a squeezing operator, and scaled rotation that provides a displacement on the canonical momentum proportional to the canonical position.
These provide the following expectation values for the canonical pair,
\begin{align}
    \ \langle \hat{q}(t) \rangle =&~  \left\{ \cos{\left[ 2 \varphi_{q}(t) \varphi_{p}(t)  \right]} \,   \langle \hat{q}(0) \rangle   + \frac{\varphi_{p}(t)}{\varphi_{q}(t)} \sin\left[  2 \varphi_{q}(t) \varphi_{p}(t) \right] \, \langle \hat{p}(0) \rangle \right\} e^{2 r(t)}, \\
    \langle \hat{p}(t) \rangle =&~ \left\{ \cos{\left[ 2 \varphi_{q}(t) \varphi_{p}(t)  \right]} e^{-2 r(t)}  - 2 \theta_{q}^{2}(t) \frac{\varphi_{p}(t)}{\varphi_{q}(t)} \sin {\left[ 2 \varphi_{q}(t) \varphi_{p}(t)  \right]} \, e^{2 r(t)} \right\} \langle \hat{p}(0) \rangle + \nonumber \\ 
    &~ -  \left\{ \frac{\varphi_{q}(t)}{\varphi_{p}(t)}   \sin\left[  2 \varphi_{q}(t) \varphi_{p}(t) \right] e^{-2 r(t)} + 2 \theta_{q}^{2}(t) \cos{\left[ 2 \varphi_{q}(t) \varphi_{p}(t)  \right]} \, e^{2 r(t)} \right\} \langle \hat{q}(0) \rangle,
\end{align}
that suggest the aforementioned rotations but now two types of scaling, an exponential scaling from the squeezing and a displacement in the canonical momentum with a scaling proportional to $\theta_{q}^{2}(t)$ from the second rotation, Fig. \ref{fig:Figure3}.
An adequate selection of time-dependent trap frequency or external driving provides us with displaced states, squeezed states, displaced squeezed states, or squeezed displaced states \cite{Moller1996p5378}.

Ideally, it is possible to engineer a time-dependent frequency or external driving that undoes any of the processes mentioned before. 
For example, if we require the state of our system at some final time $t_{f}$ to be the same as that at some initial time $t_{0}$, we can find conditions for the auxiliary time-dependent parameter functions and their time derivatives delivering the functions $\omega(t)$ and $\Omega(t)$ that undo changes in the state of the system up to an overall phase. 
Here, we aim for time-dependent resonant frequency $\omega(t)$ and driving parameter $\Omega(t)$ that yield symmetric behavior, with respect to the central time of the protocol, for the expectation values of the canonical pair and their standard deviation.
In the following, we consider the ground state of the harmonic oscillator for parameter values at time $t=0$ as the initial state, 
\begin{align}
\langle q \vert \psi(0) \rangle = \left[  \frac{ m \omega(0)}{\pi \hbar} \right]^{1/4} e^{- \frac{m \omega(0)}{2 \hbar} q^{2}},
\end{align}
and show that the analysis is valid for slow or fast changes.
For the sake of simplicity, we will use the following pulse function \cite{Ventura2019}, 
\begin{align}
    \Theta(t_{i}, t_{0}, \epsilon) =&~ \frac{1}{4} \left\{ 1 + \mathrm{erf}\left[ \epsilon \left(t - t_{i} - \frac{3}{ \epsilon}\right) \right] \right\}  \left\{ 1 + \mathrm{erf}\left[ -\epsilon \left( t - t_{o} + \frac{3}{\epsilon} \right) \right] \right\},
\end{align}
given in terms of Gauss error function $\mathrm{erf}(t) = \left( 2 / \sqrt{\pi} \right) \int_{0}^{t} e^{- \tau^{2}  d\tau }$. 
This function provides us with a step pulse starting at time $t_{i}$ and ending at time $t_{o}$ with ramp steepness controlled by the parameter $\epsilon$, a larger parameter value leads to a steeper ramp, the middle height of the ramps occurs at $ t_{i} + 3 / \epsilon$ and $ t_{o} - 3 / \epsilon$, in that order.

\begin{figure}
    \centering
    \includegraphics{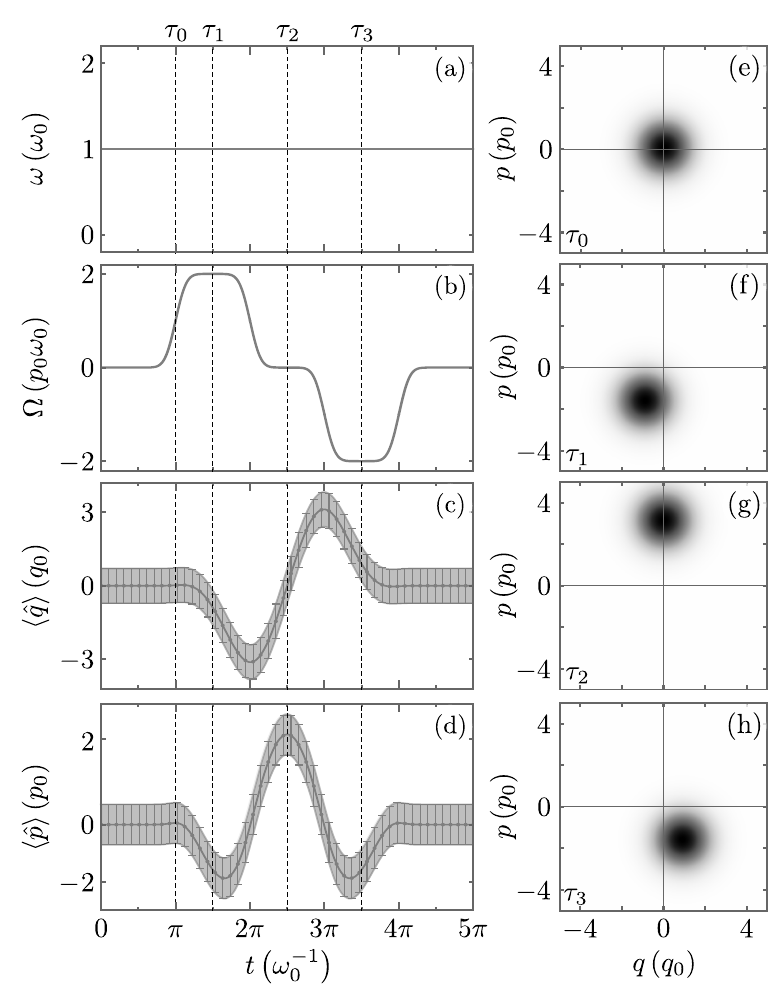}
    \caption{Slow rate of change protocol to create and remove displacement using (a) a constant frequency $\omega(t) = \omega_{0}$ and (b) a time-dependent driving strength $\Omega(t)$. 
    Solid lines show the analytic time evolution of the canonical (c) position and (d) momentum with their standard deviations as shadow regions; whisker boxes show numerical results. 
    Husimi Q-function at dimensionless time (e) $ \omega \tau _{0} = \pi $, (f) $ \omega \tau _{1} = 3 \pi / 2 $, (g) $ \omega \tau _{2} = 5 \pi / 2 $, (h) $ \omega \tau _{3} = 7 \pi / 2 $. }
    \label{fig:Figure4}
\end{figure}
\begin{figure}
    \centering
    \includegraphics{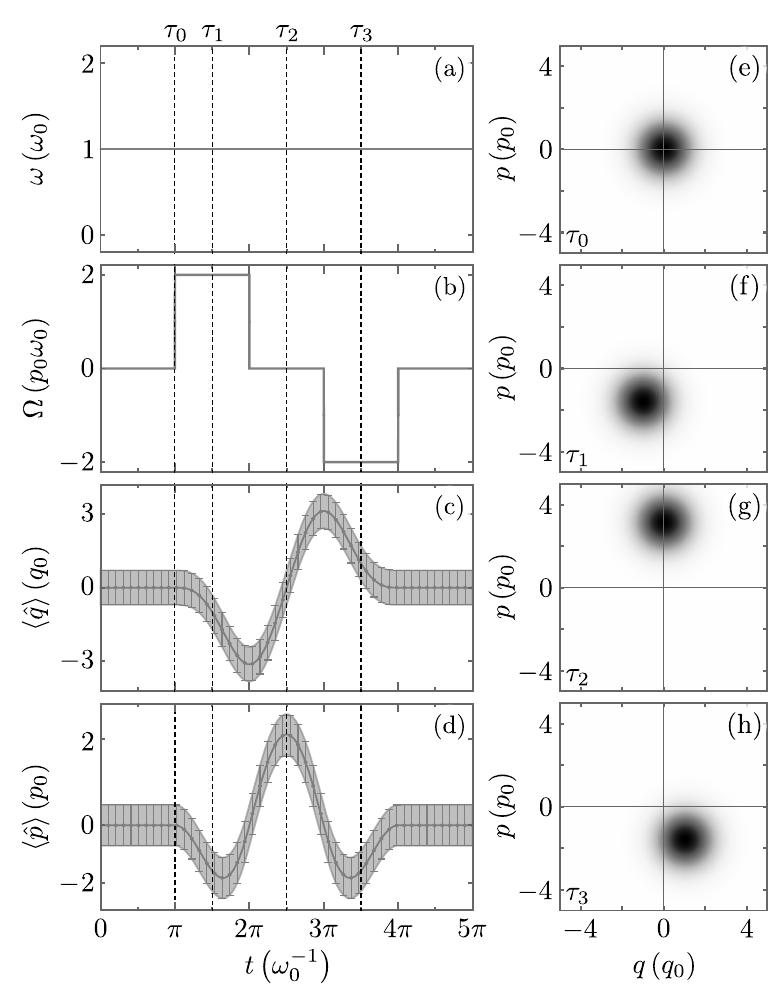}
    \caption{Same as Fig. \ref{fig:Figure4} with a fast rate of change protocol.}
    \label{fig:Figure5}
\end{figure}

First, let us consider a displacement protocol defined by a constant frequency $\omega(t) = \omega_{0}$, Fig. \ref{fig:Figure4}(a) [Fig. \ref{fig:Figure5}(a)] and a time-dependent external driving,
\begin{equation}
    \Omega(t) = 2 \left[ \Theta (t_{0}, t_{1} , \epsilon) - \Theta(t_{2}, t_{3}, \epsilon) \right] p_{0} \omega_{0},
\end{equation}
composed by two pulses starting at $t_{0} = \pi - 3\epsilon$ and $t_{2} = 3\pi - 3\epsilon$ and ending at $t_{1} = 2 \pi + 3\epsilon$ and $t_{3} = 4 \pi + 3\epsilon$, in that order, such that the middle points of their ramps are located at $j \pi$ with $j = 1, 2, 3, 4$. 
Hereby, we will use a steepness parameter $\epsilon = 2$ [$\epsilon = 1000$] for the pulses with the slow [fast] rate of change, Fig. \ref{fig:Figure4}(b)  [Fig. \ref{fig:Figure5}(b)]. 
We consider resonant driving $\omega_{d} = \omega_{0}$ and an initial driving phase $\phi = \pi/2$. 
We may divide the protocol into three legs, the first one from $t_{0}$ to $t_{1}$ covers the first pulse providing a displaced vacuum state.
The second leg, from $t_{1}$ to $t_{2}$ with constant null driving strength, only rotates the state of the system in phase space
The last leg, from $t_{2}$ to $t_{3}$, covers the second pulse that undoes the displacement generated by the first one, Fig. \ref{fig:Figure4}(c)  [Fig. \ref{fig:Figure5}(c)] and Fig. \ref{fig:Figure4}(d)  [Fig. \ref{fig:Figure5}(d)]. 
In order to undo the driving in this scenario, the time-dependent driving strength times the harmonic driving function must be an odd function with respect to the middle of the whole sequence.
We show Husimi Q-function for intermediate times of the protocol in Fig. \ref{fig:Figure4}(e) [Fig. \ref{fig:Figure5}(e)] to Fig. \ref{fig:Figure4}(h)  [Fig. \ref{fig:Figure5}(h)].
\begin{figure}
    \centering
    \includegraphics{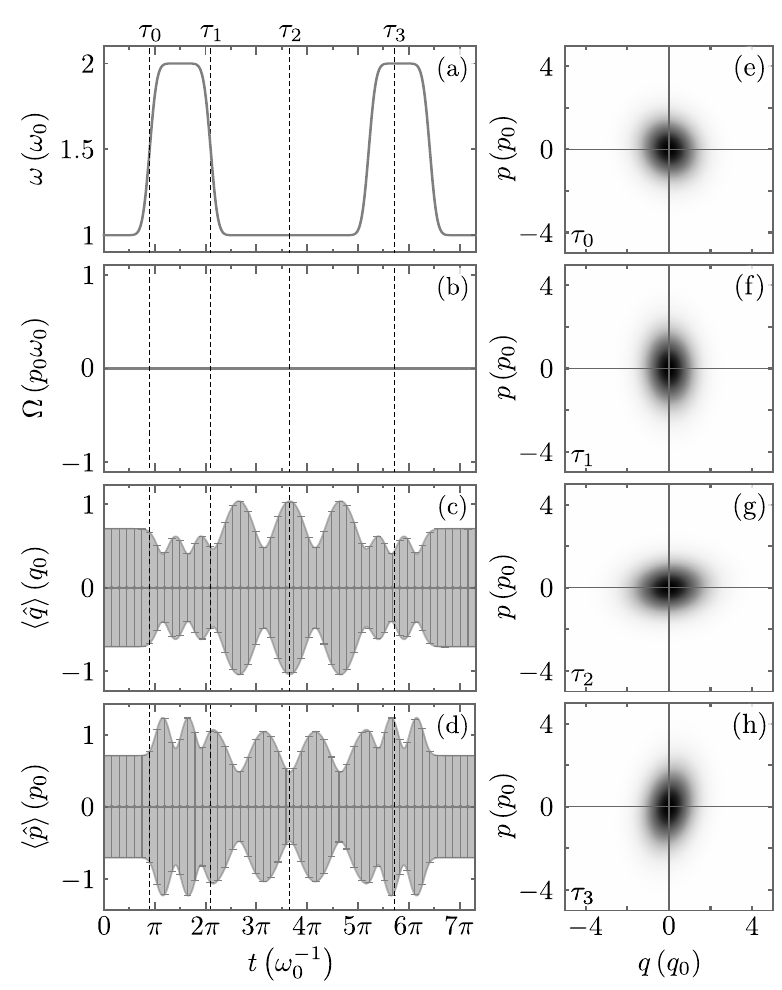}
    \caption{Slow rate of change protocol to create and remove squeezing using (a) a time-dependent frequency $\omega(t)$ and (b) a constant driving strength $\Omega(t)$. 
    Solid lines show the analytic time evolution of the canonical (c) position and (d) momentum with their standard deviations as shadow regions; whisker boxes show numerical results. 
    Husimi Q-function at dimensionless time (e) $\omega_{0} \tau_{0} = 7 \pi / 8 $, (f) $\omega_{0} \tau_{1} = 17 \pi / 8  $, (g) $\omega_{0} \tau_{2} = 11.388 $, (h) $\omega_{0} \tau_{3} = 18.064 $.}
    \label{fig:Figure6}
\end{figure}
Second, let us consider a squeezing protocol defined by a time-dependent frequency, 
\begin{equation}
    \omega(t) = \left[ 1 + \Theta (t_{0}, t_{1} , \epsilon) + \Theta(t_{2}, t_{3}, \epsilon) \right] \omega_{0},
\end{equation}
composed by two pulses starting at $t_{0} = 9 \pi / 10 - 3 / \epsilon$ [$t_{0} = 7 \pi / 8 - 3 / \epsilon$] and $t_{2} = 2 t_{m} - 21 \pi / 10 -3 \epsilon$  [$t_{2} = 2 t_{m} - 17 \pi / 8 - 3 / \epsilon$] and ending at $t_{1} = 21 \pi / 10 + 3/\epsilon$ [$t_{1}= 17 \pi /8 $] and $t_{3} = 2 t_{m} - 9 \pi /10 + 3 / \epsilon$  [$t_{3} = 2 t_{m} - 7 \pi / 8 + 3 \epsilon$] in that order, where the auxiliary time $t_{m} = 11.485$ $[t_{m} = 11.388]$ is the middle of the whole pulse sequence, Fig. \ref{fig:Figure6}(a) [Fig. \ref{fig:Figure7}(a)]. 
In absence of driving, Fig. \ref{fig:Figure6}(b) [Fig. \ref{fig:Figure7}(b)], we may divide the protocol into three legs.
The first one, from $t_{0}$ to $t_{1}$, covers the first pulse providing a squeezed vacuum state;
we optimize its duration by requiring $\theta_{q}(t_{m}) = 0$ to produce squeezing without displacement in the canonical momentum. 
The second leg, from $t_{1}$ to $t_{2}$ with constant frequency, only rotates the state of the system in phase space.
The last leg, from $t_{2}$ to $t_{3}$, covers the second pulse that undoes the squeezing generated by the first one, Fig. \ref{fig:Figure6}(c)  [Fig. \ref{fig:Figure7}(c)] and Fig. \ref{fig:Figure6}(d)  [Fig. \ref{fig:Figure7}(d)].
The key to undoing the squeezing with two symmetrical pulses is to optimize the separation between the first and second pulses such that the time-dependent frequency $\omega(t)$ makes the first derivative of the Ermakov auxiliary function zero at the middle, start and end of the protocol, $\dot{\rho}(t_{0}) = \dot{\rho}(t_{m}) = \dot{\rho}(t_{3}) = 0$.
We show Husimi Q-function for intermediate times of the protocol in Fig. \ref{fig:Figure6}(e) to Fig. \ref{fig:Figure7}(h) [Fig. \ref{fig:Figure6}(e) to Fig. \ref{fig:Figure7}(h)].
We want to note that optimizing the duration of the initial pulse by requiring both $\rho(t_{0})=\rho(t_{1})$  and $\dot{\rho}(t_{0}) = \dot{\rho}(t_{1}) = 0$ yields a process when the initial state is squeezed and returned to its initial state in a single pulse.
\begin{figure}
    \centering
    \includegraphics{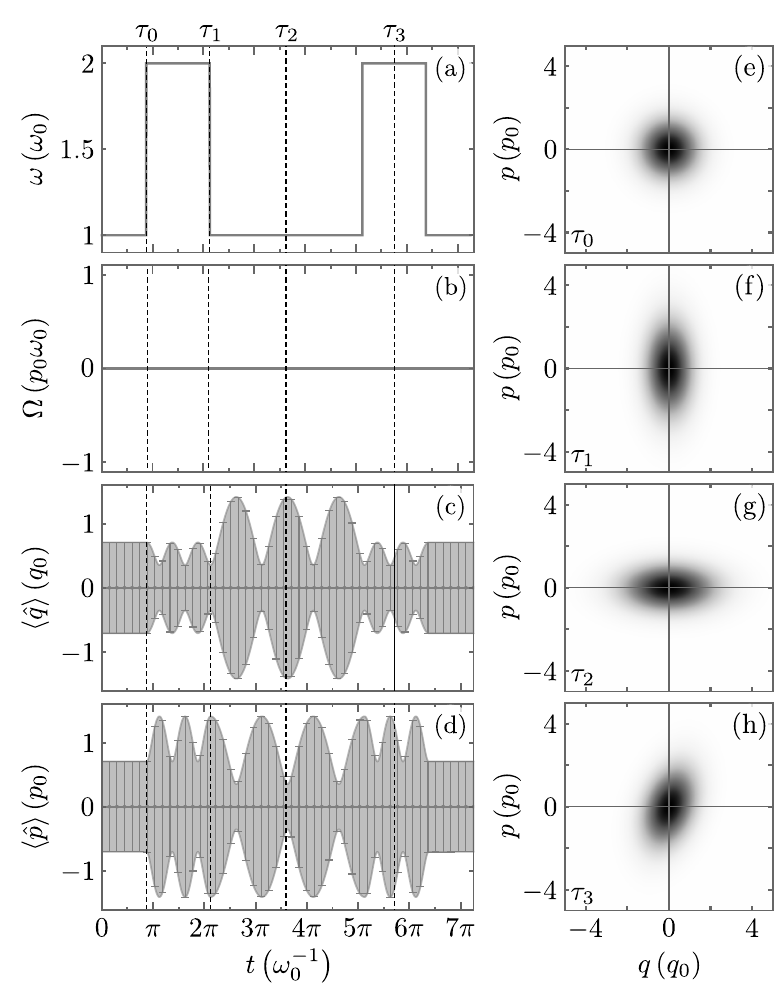}
    \caption{Same as Fig. \ref{fig:Figure6} with a fast rate of change protocol.}
    \label{fig:Figure7}
\end{figure}
\begin{figure}
    \centering
    \includegraphics{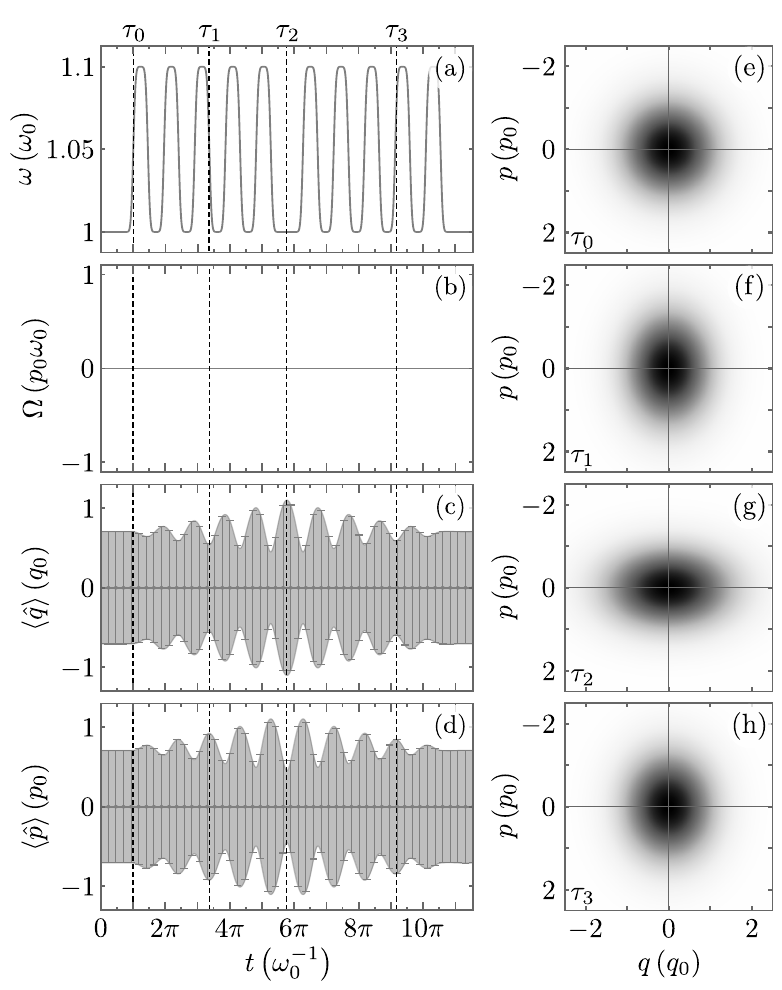}
    \caption{Slow rate of change protocol to create and remove squeezing using (a) a time-dependent frequency $\omega(t)$ and (b) a constant driving strength $\Omega(t)$. 
    Solid lines show the analytic time evolution of the canonical (c) position and (d) momentum with their standard deviations as shadow regions; whisker boxes show numerical results. 
    Husimi Q-function at dimensionless time (e) $\omega_{0} \tau_{0} = \pi$, (f) $\omega_{0} \tau_{1} = 3.363\pi$, (g) $\omega_{0} \tau_{2} = 5.772\pi $, (f) $\omega_{0} \tau_{3} = 9.136 \pi $.}
    \label{fig:Figure8}
\end{figure}
It is possible to suggest more complex protocols, for example, let us generate and, then, undo squeezing with an optimized time-dependent sequence of resonant frequency pulses,  
\begin{equation}
    \omega(t) = \left[1 + 0.1 \sum_{j=0}^{9} \Theta (t_{2j}, t_{2j+1} , \epsilon) \right] \omega_{0},
\end{equation}
where each pulse in the first half of the sequence provides squeezing by optimizing its length such that the auxiliary Ermakov parameter $\rho(t)$ has local maxima at the start of the pulses $t_{2j}$ and local minima at their end $t_{2j+1}$,  Fig. \ref{fig:Figure8}(a) [Fig. \ref{fig:Figure9}(a)], without external driving, Fig. \ref{fig:Figure8}(b) [Fig. \ref{fig:Figure9}(b)].
Under these conditions, each of the pulses in the first half of the sequence will increase the squeezing in the state of the system one at a time.
Again, the key to undoing the squeezing with two symmetrical pulse trains is to optimize the separation between the first and second pulse trains by requiring that the first derivative of the Ermakov auxiliary function becomes null at the start, middle, and end time of the protocol, $\dot{\rho}(t_{0}) = \dot{\rho}(t_{m}) = \dot{\rho}(t_{20}) = 0$.
This will bring back the system to its initial state, Fig. \ref{fig:Figure8}(c)  [Fig. \ref{fig:Figure9}(c)] and Fig. \ref{fig:Figure8}(d)  [Fig. \ref{fig:Figure9}(d)], up to an overall phase.
We show Husimi Q-function for intermediate times of the protocol in Fig. \ref{fig:Figure8}(e) to Fig. \ref{fig:Figure9}(h) [Fig. \ref{fig:Figure8}(e) to Fig. \ref{fig:Figure9}(h)].
We want to stress that in this example, the difference between minimum and maximum resonant frequency is ten percent of the original resonant frequency and produces notable squeezing while the difference in the preceding protocol is a hundred percent of the original resonant frequency.
\begin{figure}
    \centering
    \includegraphics{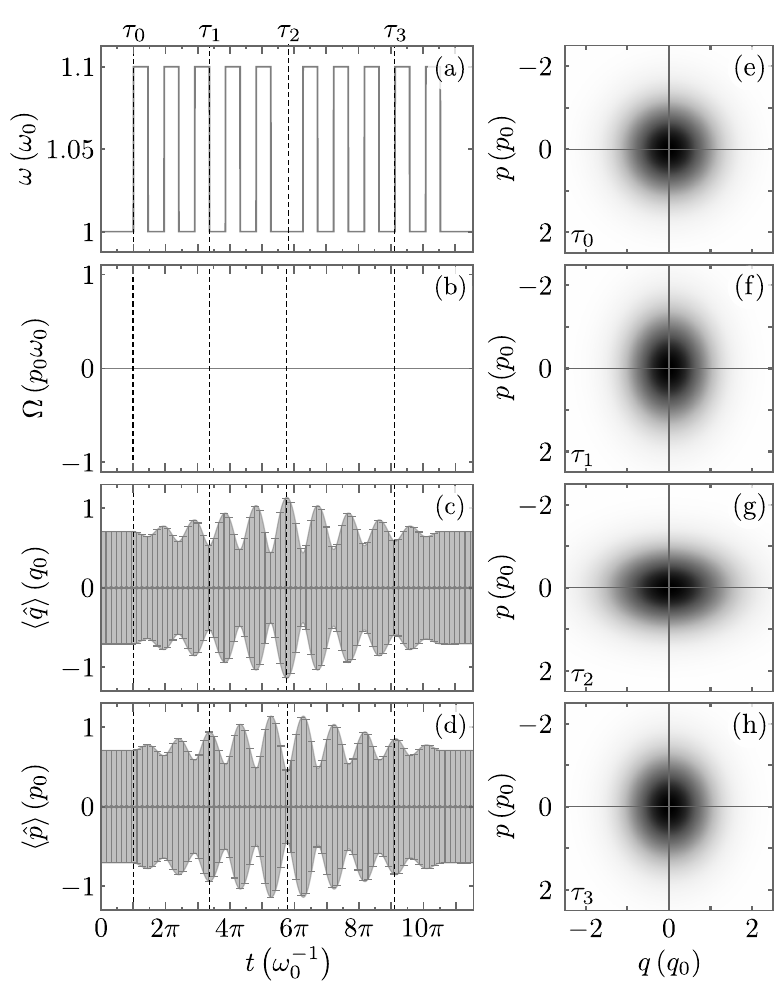}
    \caption{Same as Fig. \ref{fig:Figure8} with a fast rate of change protocol.}
    \label{fig:Figure9}
\end{figure}
%

\section{Conclusion} \label{sec:Conclusion}

We presented a Lie algebraic approach to the time-dependent driven quantum harmonic oscillator.
It yields an analytic form for the final state of the system in terms of the three basic operations provided by the action of the group on the algebra, that is rotation, displacement, and squeezing, and a set of differential equations for their parameters.

Our approach works for slow or fast changes of the time-dependent quantum harmonic oscillator parameters as long as they are continuous and differentiable. 
Its intuitive form, in terms of the three basic operations, allows us to propose protocols for state engineering that range from the obvious generation of displaced or squeezed states to more complex states.
We show protocols to undo displacement and squeezing in closed systems that may provide a reference for the characterization of open systems starting in the ground state of the initial time of the system. 
Another straightforward proposal is the optimization of squeezing, for example, via a time-dependent frequency that takes combines a pulse with free propagation that may be useful for systems whose change in resonant frequency is limited.

Our Lie algebraic approach is an example of so-called shortcuts to adiabaticity and may provide an analytical framework for the optimization of quantum state engineering or the training of machine learning schemes for the control of quantum systems.


%

\end{document}